# A novel approach to Service Discovery in Mobile Adhoc Network

*Nomman Islam*
Center of Research in Ubiquitous Computing
National University of Computer & Emerging Sciences
Karachi, Pakistan

*Zubair A.. Shaikh*
Center of Research in Ubiquitous Computing
National University of Computer & Emerging Sciences
Karachi, Pakistan

*Abstract*—Mobile Adhoc Network (MANET) is a network of a number of mobile routers and associated hosts, organized in a random fashion via wireless links [16]. During recent years MANET has gained enormous amount of attention and has been widely used for not only military purposes but for search-and-rescue operations, intelligent transportation system, data collection, virtual classrooms and ubiquitous computing [14]. Service Discovery is one of the most important issues in MANET. It is defined as the process of facilitating service providers to advertise their services in a dynamic way and to allow consumers to discover and access those services in an efficient and scalable manner [20]. In this paper, we are proposing a flexible and efficient approach to service discovery for MANET by extending the work of [5]. Most of the service discovery protocols proposed in literature don't provide an appropriate route from consumer to service provider. Hence after services are discovered, a route request needs to be initiated in order to access the service. In this paper, we are proposing a robust and flexible approach to service discovery for MANET that not only discovers a service provider in the vicinity of a node, but at the same time, it also provides a route to access the service. We have extended the approach proposed in [5] for efficiency by adding the push capabilities to service discovery

## I. INTRODUCTION

A mobile ad hoc network (MANET) is a self-configuring network that is formed and deformed on the fly by a collection of mobile nodes without the help of any prior infra-structure or centralized management [14]. These networks are characterized as infrastructure less, mobile, autonomous, multi-hopped, self-organized and self-administered, having dynamic topology and unpredictable traffic patterns etc.

A great deal of research is being carried out to solve various issues of MANET. These issues include Routing, MAC Layer Issues, Power Management, Transport protocol, Quality of Service, Billing, Addressing, Service Discovery, Data Management and Security etc [11], [13] and [15].

One of the most important questions in MANET is the discovery of services available around the vicinity of any node. A service can be any hardware, software or any other entity that a user might be interested to utilize and Service Discovery is the process of discovering the services based on user preferences. An efficient and scalable approach to service discovery can lead to the development a large number of potential applications. For example, in a vehicular ad hoc network, vehicles might be interested in knowing the services provided by a near by fuel station. Similarly, in a battlefield soldiers might be interested in sharing the situation about the whole battlefield.

Due to the dynamic nature of MANET, there are always spatial and temporal variations in the availability of services [18]. Hence a service discovery strategy should be highly robust, efficient and dynamic in nature.

In this paper we are proposing an efficient and flexible approach to service discovery in MANET by extending the work of [5]. Rest of the sections is organized as follows: first we will provide a review of existing approaches to service discovery. We will then discuss our approach to Service Discovery along with implementation details. Finally we will conclude the paper with future work.

## II. LITERATURE REVIEW

The service discovery protocols that have been proposed in the literature can be classified as directory less (e.g. IBM DEAPspace, UPnP and Konark) or directory based architecture (e.g. Salutation, JINI, SLP) [19]. SLP [20],[9] and JINI[20],[10] relies on a directory to store services. SLP represents the service by means of URL and attributes. JINI is based on Java and uses Interfaces and RMI mechanisms for service discovery process. Because of the lack of semantic information, these protocols can only perform exact matching of services using identifier and attributes.

To support service discovery based on semantics, DReggie[8] (an extension of JINI) is proposed. DReggie semantically represents the services by means of DAML and compares the service using a PROLOG reasoning engine. GSD[3] also describes the services semantically by means of DAML. They group the services based on semantics for efficient discovery of services.

In order to address heterogeneity of MANET, [2] proposed a middleware Konark, designed specifically for discovery and delivery of device independent services in ad-hoc networks. Konark is based on SOAP and HTTP for service discovery and delivery purposes.

[5] proposed a totally different approach to service discovery by extending AODV protocol and provided service

discovery at network layer. As a result when a service discovery request is initiated to discover a service, a route is also established towards the service provider. Hence, when the client wants to use the service, a new route request is not required.

### III. A NOVEL APPROACH TO SERVICE DISCOVERY IN MANET

We are proposing an efficient approach to service discovery for MANET by extending the concept proposed in [5]. Nodes will be using an extension of AODV routing protocol to search on demand for a service provider along with an appropriate route to access that service. We are extending the concept of [5] by allowing a node to not only pull the service provider information on-demand, but a node will also be pushing the service advertisements periodically along with the route information.

The novelty of our strategy is that unlike most of the service discovery protocols (e.g. JINI, UPnP, Deapspace and SLP), an appropriate path towards the service provider is available after the completion of service discovery process. In addition to employing AODV, we are also explicitly pushing the service advertisement to adjacent nodes along with appropriate routing information. This information can be saved by adjacent nodes based on their preferences and can also be propagated ahead. So, in most of the cases, a service discovery request can be served from nodes local service table and doesn't need to be propagated on the network. The integration of service discovery with the network layer is motivated by the survey done in [21]. The development of a cross layer service discovery solution is always useful in the network of energy constrained mobile devices because: 1) putting service discovery at the network layer will reduce the control messages overhead 2) less number of messages are exchanged among devices, hence processing overhead is reduced. We believe that the periodic advertisements of services (as done in our approach) will not cause too much traffic overhead because a number of service advertisements can always be transmitted in a single message. Also, for each service discovery, service discovery and route determination is done in a single request. Hence, less number of messages is exchanged among nodes. The disadvantage of our approach however is that it is dependent on routing protocol and can work only for AODV routing protocol.

Fig 3 describes the details of our proposed approach in pseudo code. Every node will maintain a Routing Table to hold routing information and a Service table to maintain information about the services provided by it and other nodes along with other information like the life time of the services, the provider of the service etc. The routing and service table maintained by every node is shown in Fig 2.

The nodes will periodically broadcast the services provided by it to its peer nodes by means of a message UPDATE_SERVICE_TABLE (UST). The message contains the service advertisements along with the routing information corresponding to the provider of each service.

The receiving node can store the service advertisements it heard from its neighbor based on its preferences and interests.

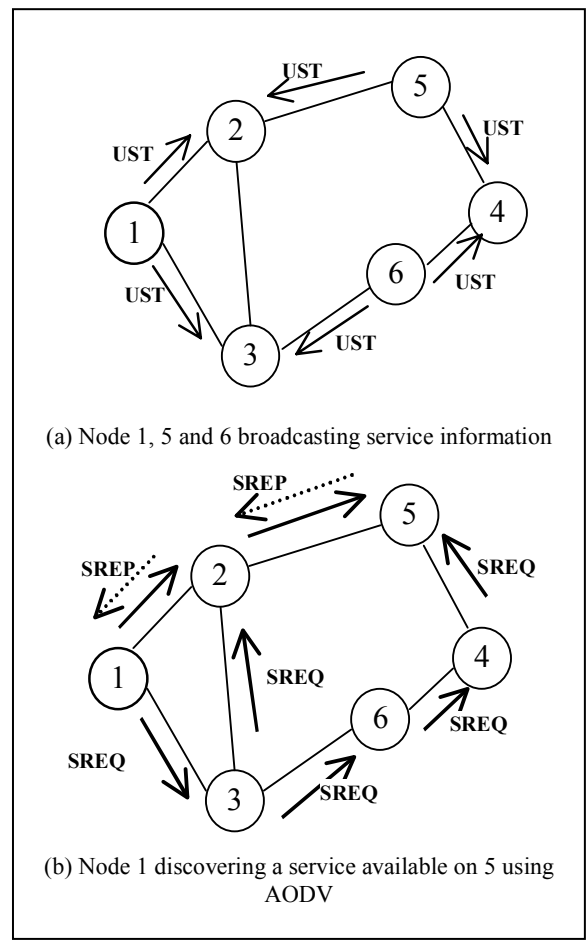

(a) Node 1, 5 and 6 broadcasting service information

(b) Node 1 discovering a service available on 5 using AODV

Fig 1: A Novel approach to Service Discovery

In addition, to avoid network congestion, if a node hears a service advertisement it reschedules its broadcast at a later time to avoid network congestion. If a node wants a service, it will first of all check its service table for a possible match. If a service is found in the service table, node can immediately requests the provider for the desired service. However, if no appropriate service provider is found in the local service table, a service discovery request 'SREQ' will be issued to find out a service provider for this service. The SREQ is served based on 'Ad hoc On Demand Distance Vector Routing Protocol' [5]. The SREQ message is propagated along the network. Any node receiving the SREQ message will first check in its service table for corresponding service. If a match is found, an appropriate reply 'SREP' message will be generated. If no match is found, the intermediate nodes will propagate the message SREQ to adjacent neighbors. In addition to that, it will create a temporary reverse route towards source of the message. When a node receives an SREP message, it will check its routing table, delete the corresponding temporary reverse route and create a forward entry towards the destination. The node will then send the SREP message towards the source. The RREP message will contain complete details of the services it provides.

In the routing table, along with each valid route a list of precursors is also maintained. The precursor lists are the set of

(a) Routing Table

| | |
|---|---|
| SequenceNumber | Determines the freshness of the routing information |
| Destination | The destination to be reached |
| HopCount | The number of hop counts |
| NextNode | The next node to whom packet will be forwarded |
| Status | status of the route (temporary or a permanent route) |
| Precursors | The upstream nodes of a particular route |

(b) Service Table

| | |
|---|---|
| Expiration Time | service expiration time |
| Provider | service provider name |
| Service Name | name of the service |
| Service Type | the type of the service |
| Service Description | details of service or a URL where details can be found |

Fig 2 – Routing and Service Tables maintained by nodes

nodes that may be forwarding packets on this route. In case of link breakage or a node failure, an RERR message will be generated to notify the precursors immediately about the link breakage, which in turn notify its upstream nodes and so on until source node reaches. If no precursor list is available, the 'RERR' message will be broadcasted along the network. In order to avoid network congestion a broadcasted message will have a unique id and a node will not broadcast RERR message with same id twice. Fig 1 shows this over all process in detail.

## IV. SIMULATION RESULTS

To simulate our approach, we have developed our own simulation software. The simulator creates 'n' nodes dynamically and links them randomly. Every node is randomly chosen to be either capable of storing or not storing advertisements. Nodes are randomly assigned a number of services. While simulation is running, after every half second, a random number of requests are generated and based on a probability of 0.5 either a link is randomly created or broken. Fig 4 shows the result of our simulation. We have compared the results based on Request Acquisition Latency i.e. the time duration in which a service discovery request is initiated and the service discovery process is completed. Fig 4 compares the result when no service advertisement is broadcasted periodically with the result when broadcast capability is added. Fig 4(a) and 4(b) shows the situation when 50 nodes were simulated with a total of 25 services available on the network. In this situation, 174 requests were generated in a simulation time of 30 seconds. When no broadcasting was done 147 requests were replied with request acquisition latency between 0-0.2s. But when broadcasting was done, the 154 requests were

```
While Node is Running

  If 6 seconds have elapsed since last broadcast
    deleteTemporaryRoutes();
    cleanUpCache();
    cleanUpServiceTable();
    braodCastRoutingNServiceEntriesToNeighbours();
  End If;

  Packet = Read_A_Message();

  If PacketType = UPDATE_SERVICE_TABLE
    If NodeCanStoreAdvertisement
      storeServiceAdvertisementGot();
      storeRoutingInfoGot();
    End If;
  End If;

  If PacketType = SREQ
    If This SREQ request has already been entertained
    {Duplicate SREQ might arrive from different paths}
      Return;
    Else If SREQ can be served by a lookup on local
    service table
      send SREP back to the source of SREQ
    Else
      BroadCast SREQ to adjacent nodes
      Add a reverse routing  entry towards source of
      SREQ with next node set to the source hop of
      SREQ
    End If;  {End Else}
  End If;

  If PacketType = SREP
    If this is the reply of my SREQ request then
      create a forward routing entry towards the source
      of SREP and set next node the source_hop of
      SREP
    Else
      Remove corresponding temporary routes from
      Routing table
      Create a forward entry towards the source of
      SREP with next node set to source_hop of SREP
      Forward SREP to my upstream node
    End If;  {End Else}
  End If;

  If PacketType = RERR
    If RouteDeleted Has PreCursors
      If I had any route towards the now un-reachable
      Node
        informPrecursorsAboutTheRouteDeletion();
        removeCorrspondingRouteFromRoutingTable();
      End If;
    Else  If I had any route towards the now un-
    reachable node
      broadCastToNeighborsAboutRouteDeletion();
      removeCorrspondingRouteFromRoutingTable();
    End If;           {End Else}
  End;  {End RERR If}

End While
```

Fig 3 – Pseudo Code Describing the Service Discovery Procedure

replied in between 0-0.2s. This shows that by adding broadcasting functionality, performance of service discovery strategy has improved. Similarly in the case of 100 nodes, we

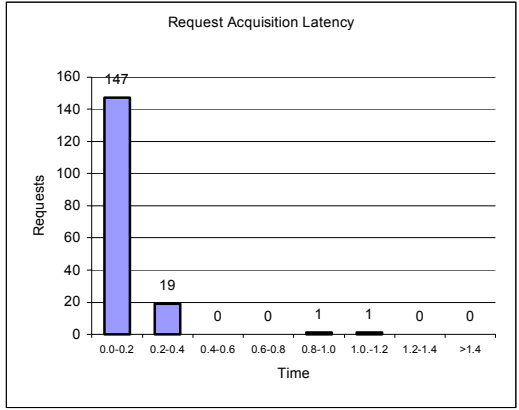
a) Without Broadcast

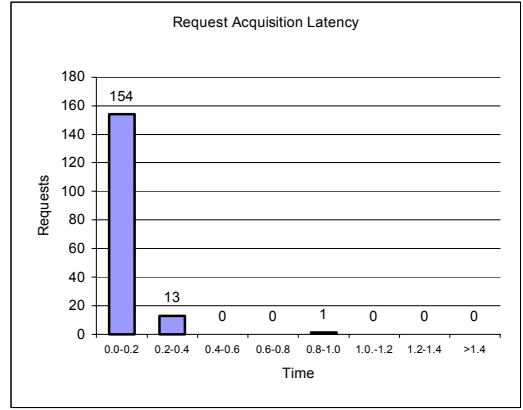
b) With Broadcast

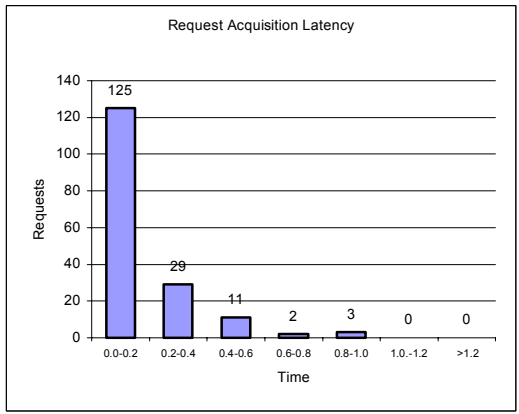
c) Without Broadcast

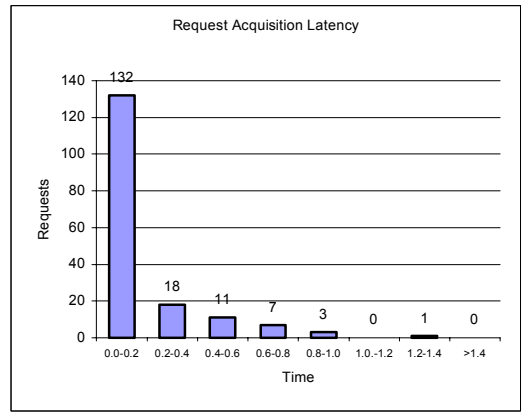
d) With Broadcast

Fig 4 – Comparison of Request Acquisition Latency when no Broadcasting done with the result when Broadcast capability is added

can see in Fig 4(c) and 4(d) that broadcasting of service advertisements has improved the request acquisition latency.

## V. CONCLUSION

In this paper we have proposed a novel approach towards service discovery in MANET. The proposed approach is very flexible, efficient and can be adopted to work in any environment. The current work can be extended to represent the service using better representation language like DAML etc. In addition broadcasting of service advertisement is right now done periodically. We can improve the broadcasting mechanism. The broadcasting can be a function of network congestion, service popularity factor etc.

## ACKNOWLEDGMENT

This research work is supported by 'Center of Research in Ubiquitous Computing', National University of Computer and Emerging Sciences, Karachi, Pakistan